\documentclass{aa}
\usepackage{txfonts}
\usepackage[authoryear]{natbib}
\usepackage{natbib}
\usepackage{amssymb}
\bibpunct{(}{)}{;}{a}{}{,}
\usepackage{epsf}
\usepackage{graphicx}
\usepackage{epsfig}
\usepackage{graphics}
\usepackage{subfigure}
\usepackage[english]{babel}
\usepackage{rotating}

\newcommand{\teff}{$T_{\rm{eff}}$}
\newcommand{\logg}{$\log g$}
\newcommand{\lL}{\ifmmode \log \frac{L}{L_{\sun}} \else $\log \frac{L}{L_{\sun
}}$\fi}
\newcommand{\mdot}{$\dot{M}$}
\newcommand{\myr}{M$_{\sun}$ yr$^{-1}$}
\newcommand{\vsini}{$V$~sin$i$}
\newcommand{\vinf}{$v_{\infty}$}

\newcommand{\vmac}{$v_{\rm mac}$}

\newcommand{\kms}{km s$^{-1}$}
\newcommand{\msun}{M$_{\sun}$}
\newcommand{\rsun}{R$_{\sun}$}
\newcommand{\tspin}{$\tau_{\rm{spin}}$}

\begin{document}

\title{Observational effects of magnetism in O stars: surface nitrogen abundances \thanks{Based on observations collected at the CFHT and the T\'elescope Bernard Lyot.}}
\author{F. Martins\inst{1}
\and C. Escolano\inst{1,4} 
\and G.A. Wade\inst{2}
\and J.~F. Donati\inst{3}
\and J.~C. Bouret\inst{4,5}
\and the MiMeS collaboration
}
\institute{LUPM--UMR 5299, CNRS \& Universit\'e Montpellier II, Place Eug\`ene Bataillon, F-34095, Montpellier Cedex 05, France \\
           \email{fabrice.martins@univ-montp2.fr}
           \and
           Dept. of Physics, Royal Military College of Canada, PO Box 17000, Stn Forces, Kingston, Ontario K7K 7B4, Canada  
           \and 
           IRAP--UMR 5277, CNRS \& Universit\'e Paul Sabatier, 16 Avenue Edouard Belin, F-31400, Toulouse, France 
           \and
           LAM--UMR 6110, CNRS \& Universit\'e de Provence, rue Fr\'ed\'eric Joliot-Curie, F-13388, Marseille Cedex 13, France 
         \and
             NASA/GSFC, Code 665, Greenbelt, MD 20771, USA \\
}

\date{Recieved September 7$^{\rm th}$ 2011 / Accepted 9$^{\rm th}$ December 2011}

\abstract
{}
{We investigate the surface nitrogen content of the six magnetic O stars known to date as well as of the early B--type star $\tau$ Sco. We compare these abundances to predictions of evolutionary models to isolate the effects of magnetic field on the transport of elements in stellar interiors.}
{We conduct a quantitative spectroscopic analysis of the sample stars with state-of-the-art atmosphere models. We rely on high signal-to-noise ratio, high resolution optical spectra obtained with ESPADONS at CFHT and NARVAL at TBL. Atmosphere models and synthetic spectra are computed with the code CMFGEN. Values of $N/H$ together with their uncertainties are determined and compared to predictions of evolutionary models.}
{We find that the magnetic stars can be divided into two groups: one with stars displaying no N enrichment (one object); and one with stars most likely showing extra N enrichment (5 objects). For one star ($\Theta^{1}$ Ori C) no robust conclusion can be drawn due to its young age. The star with no N enrichment is the one with the weakest magnetic field, possibly of dynamo origin. It might be a star having experienced strong magnetic braking under the condition of solid body rotation, but its rotational velocity is still relatively large. The five stars with high N content were probably slow rotators on the zero age main sequence, but they have surface $N/H$ typical of normal O stars, indicating that the presence of a (probably fossil) magnetic field leads to extra enrichment. These stars may have a strong differential rotation inducing shear mixing. Our results should be viewed as a basis on which new theoretical simulations can rely to better understand the effect of magnetism on the evolution of massive stars.}
{}

\keywords{Stars: massive -- Stars: atmospheres -- Stars: fundamental parameters -- Stars: abundances -- Stars: magnetic field}

\authorrunning{F. Martins et al.}
\titlerunning{Nitrogen abundance of magnetic O stars}

\maketitle

%%%%%%%%%%%%%%%%%%%%%%%%%%%%%%%%%%%%%%%%%%%%%%%%%%%%%%%%%%%%%%%%%%%%%%%%%%%%%%%%%%%%%%%%%%%%%%%%%%%%%%%%%%%%%%%%%%%%%%%%%%%%%%%
%%%%%%%%%%%%%%%%%%%%%%%%%%%%%%%%%%%%%%%%%%%%%%%%%%%%%%%%%%%%%%%%%%%%%%%%%%%%%%%%%%%%%%%%%%%%%%%%%%%%%%%%%%%%%%%%%%%%%%%%%%%%%%%
\section{Introduction}

The evolution of massive stars is governed by two main physical processes: their mass loss \citep{cm86} and their rotation \citep{mm00}.  All additional processes directly affecting these properties are thus major contributors to the fate of massive stars. This is the case of magnetism. Since the pioneering work of \citet{bm97}, it is known that the presence of a large scale surface magnetic field will modify massive stars winds. By deflecting material along the field lines, magnetism affects the wind geometry and structure. Simulations by \citet{ud02,ud08,ud09} have shown that equatorial overdensities can be created. Material can either fall back onto the stellar surface or be ejected in the magnetic equatorial plane depending on the interplay with rotation. In the case of very strong magnetic fields, observations and simulations show that a rigid magnetosphere strongly affects the wind structure, creating caps of stellar material in the minima of the magneto-rotational potential \citep{townsend07}. Since the wind flow is deflected by the presence of a magnetic field, the way the star loses its angular momentum is affected. In addition, the internal structure of the star is also affected, forcing the star to follow new evolutionary paths.
Given the relevance of massive star evolution to many fields of astrophysics (galactic chemical evolution, interstellar dynamics, supernovae and gamma--ray bursts, nucleosynthesis...), it is crucial to understand the magnetic properties of these objects in order to constrain their influence on the evolution and fate of OB stars.  

The recent development of powerful spectropolarimeters has revolutionized our view of massive stars magnetism. The first detection of a magnetic field on the Trapezium O--type star $\Theta^{1}$~Ori~C \citep{donati02} dates back to less than a decade. This spectroscopically varying and strong X--ray emitting star has revealed a surface kilogauss field organized mostly as a dipole. Subsequent detections have been made on HD~191612 \citep{donati06}, HD~148937 \citep{hubrig08,wade11}, HD~57682 \citep{grunhut09} and HD~108 \citep{martins10}. Field strengths are of several hundred Gauss to a few kilogauss and the geometry, when constrained, is consistent with a dipole. The only exception is the O supergiant $\zeta$~Ori~A \citep{bouret08} for which a more complex field topology was tentatively derived, together with a remarkably weak field of 50--100 G. This raises the question of the origin of magnetism in massive stars. The debate is ongoing. A fossil origin, stable over Myrs and with a rather simple geometry, is supported by some simulations \citep{brait04,duez10}. In that case, the star retains the original magnetic field of its parental molecular cloud or dynamo field of its convective progenitor. Alternatively, dynamo processes may be at work. In analogy with low mass stars where a convective envelope (or even a fully convective structure) exists, magnetic field might be generated in the convective core of high mass stars and subsequently transported to the surface \citep{charb01,brun05}. A dynamo may also operate in the radiative zones of massive stars \citep{spruit02,mcdo04}. The field would be produced by the shear in this region. This idea has been tested in numerical simulations, with contradictory results \citep{brait06,zahn07}. 

However, assuming that this dynamo process was able to produce and maintain a magnetic field, evolutionary calculations have been conducted to test the effects of this type of magnetism on the evolution of massive stars \citep{mm04}. They find that the internal structure is strongly affected, with a completely different rotation profile in the star compared to non-magnetic calculations including rotation. A consequence of this re-organization of the stellar interior is a modification of the transport of angular momentum and of chemical elements. In particular, \citet{mm05} studied the effects on the surface chemical appearance of the star. They show that the presence of a dynamo generated magnetic field significantly increases the efficiency of mixing, leading to strong nitrogen and even helium surface enhancement. Further analysis of the feedback effects of magnetic braking balanced these results though, highlighting the importance of the type of rotation profile (differential versus solid-body) in the final chemical appearance of the star \citep{meynet11}.

Although still limited, the sample of magnetic O stars is now populous enough to test these predictions. An important side effect of the spectropolarimetric observations is the production of very high S/N ratio optical spectra at high resolution. These are perfectly suited for abundance studies. In this paper, we thus analyze the surface nitrogen content of the magnetic O stars known to date as well as of the B0.2V star $\tau$ Sco. Our aim is to see if peculiar chemical patterns are observed and if so, to what extent they can be explained by the current generation of evolutionary models. The goal is to identify specific patterns that can only be attributed to the presence of a magnetic field. This is particularly important in the context of explaining the population of N--rich slowly rotating stars in the Magellanic Cloud reported by \citet{hunter08,hunter09}. Such stars are not explained by rotating evolutionary tracks and have been suggested to be magnetic. The observations are presented in Sect.\ \ref{s_obs}. We describe the analysis with atmosphere models in Sect.\ \ref{s_mod}. The results are discussed in Sect.\ \ref{s_mag}. We show how the chemical properties of the magnetic O stars can be interpreted in terms of evolutionary models with rotation and with or without magnetic field. We finally summarize our conclusions in Sect.\ \ref{s_conc}.

%%%%%%%%%%%%%%%%%%%%%%%%%%%%%%%%%%%%%%%%%%%%%%%%%%%%%%%%%%%%%%%%%%%%%%%%%%%%%%%%%%%%%%%%%%%%%%%%%%%%%%%%%%%%%%%%%%%%%%%%%%%%%%%
%%%%%%%%%%%%%%%%%%%%%%%%%%%%%%%%%%%%%%%%%%%%%%%%%%%%%%%%%%%%%%%%%%%%%%%%%%%%%%%%%%%%%%%%%%%%%%%%%%%%%%%%%%%%%%%%%%%%%%%%%%%%%%%
\section{Stellar sample and observational data}
\label{s_obs}

%%%%%%%%%%%%%%%%%%%%%%%%%%%%%%%%%%%%%%%%%%%%%%%
\subsection{Observations}

The observations were collected with \'echelle
spectropolarimeters ESPaDOnS at CFHT (Canada--France--Hawaii
Telescope) and NARVAL at TBL (T\'elescope Bernard Lyot), partly from
the MiMeS project \citep{wade10}. The spectra cover the wavelength
range 3700--10500 \AA\ with a resolution of 68000 (ESPaDOnS) and 65000
(NARVAL). We gathered multiple sequences for each of the sample stars,
each sequence consisting of four individual subexposures taken in
different polarimeter configurations.  From these four individual
subexposures, mean Stokes parameters were derived. All frames were
processed using Libre-ESpRIT \citep{donati97b}, a fully automatic
reduction package installed at CFHT and TBL for optimal extraction of
ESPaDOnS and NARVAL spectra. The peak signal-to-noise ratios (SNRs)
per 2.6 km s$^{-1}$ velocity bin depends on the instrument, on the
exposure time and on weather conditions.  Complete references to the
spectropolarimetric data are gathered in Table~\ref{tabA1}. A summary
of the data available is given in Tab.\ \ref{tab1}. We used the
average of the entire set of spectra for each star, ensuring that no
variability in the key diagnostic lines was present among the data.

\begin{table*}[ht]
\centering
\caption{Summary of observations.}
\begin{tabular}{lccccc}
\hline
\hline
Star & Instrument & Total Exp. time & Number of spectra & typical S/N per spectrum \\
\hline
\noalign{\smallskip}
HD~108                          & NARVAL   & 34 h 20 min & 37 & 150--500 \\
HD~148937			&  ESPADONS & 4 h 00 min & 6 & 200--700 \\
HD~191612               	& ESPADONS & 2 h 40 min & 8 & 100--350 \\
$\Theta^{1}$ Ori C 		& NARVAL & 53 min & 2 & 500--1300 \\
$\zeta$ Ori A 			& NARVAL & 13 h 52 min & 336 & 1000--1400 \\
HD~57682 	                & ESPADONS & 3 h 4 min & 5 & 400--800 \\
$\tau$ Sco                      & ESPADONS & 13 min & 5 & 600--1200 \\
\hline
\end{tabular}
\tablefoot{The complete journal of observations is given in Appendix \ref{tabA1}}
\label{tab1}
\end{table*}

%%%%%%%%%%%%%%%%%%%%%%%%%%%%%%%%%%%%%%%%%%%%%%%
\subsection{Sample stars}

Our sample is composed of the six magnetic O stars known to date. Here
we give a few details about each of these objects. The journal of
observations is shown in Table \ref{tabA1}.

HD~191612 is a member of the Of?p class, first introduced by
\cite{walborn72}. Such stars exhibit, among other peculiarities,
variability in their Balmer, \ion{He}{i}, \ion{C}{iii} and
\ion{Si}{iii} lines. \citet{donati06} reported the detection of a
kilogauss magnetic field on its surface. It is presently
well-established that HD~191612 displays two recurrent states, with a
period of approximately 538 days
\citep{walborn03,walborn07,howarth07}. These two states are referenced
here as "quiet" (i.e. O8 spectral type, H$\alpha$ in absorption) and
"active" (i.e. O6 spectral type, strong wind signatures). We decided
to focus on the quiet phase since the photospheric lines, used to
derive the stellar parameters and surface abundances, are less
contaminated by wind emission than in the active phase. We relied on
data taken on August 23$^{\rm rd}$ 2009, corresponding to a phase of 0.393
\citep[see][]{wade191612}. The journal of observations in
Table\ \ref{tab1} presents additional information.

HD~108 is the second Of?p star of our sample. It is very similar to
HD~191612, with a long--term variability both photometrically and
spectroscopically. However the timescale of this variability is much
longer and is estimated to be of about 55 years. The \ion{He}{i} and
Balmer lines are the most affected, leading to a spectral type
changing from 04 to 08. \citet{martins10} reported the detection of a
(most likely) kilogauss magnetic field on its surface. We used the
average of the July-August 2009 subset of the Martins et
al.\ observations in the present study since, as for HD~191612, this
corresponds to the minimum of extra emission in the Balmer and
\ion{He}{i} lines.

HD~148937 is the last member of the Of?p class in our sample.  Its
variability is much more limited (lower amplitude) than that observed
in HD~191612 and HD~108 spectra \citep{naze08a}. The periodicity of
the variations of about 7 days, is extremely short compared to the
long periods observed in the two other Of?p stars. The HD~148937
spectra were collected during six nights in June 2010.

$\Theta^{1}$ Ori C (HD~37022) is a mid--late type O dwarf. It is the first O star
for which a magnetic field was detected \citep{donati02}. It is known
for its variability, with a period of approximately 15.4 days
\citep{stahl96, wade06}, the most striking variable spectral features
being an inverted P-Cygni profile in the \ion{He}{ii}\,$\lambda$4686
line and strong emission features in the Balmer lines (especially
H$\alpha$ and H$\beta$). The equivalent width of some He and metal
lines also vary as a function of the phase \citep[][and references
  therein]{simon06}.

Spectropolarimetric observations of $\zeta$ Ori A were collected with
NARVAL in 2007 October. Altogether, 292 circular polarization
sequences were taken. The peak SNR per 2.6 km s$^{-1}$ velocity bin
varies between 800 and 1500. $\zeta$ Ori A probably has a short
rotation period of approximately 7 days and a weak magnetic field
\citep{bouret08}. Although some spectral lines appear variable, the
main diagnostics used to derive photospheric parameters remain stable
during the rotation cycle.

HD~57682 is an O9 sub-giant. Its magnetic field of strength 1.7 kG was
reported by \citet{grunhut09}. HD~57682 displays spectral variability but on a much
lower level compared to HD~191612 and HD~108. Its rotational period is
estimated to be $\sim$64 days \citep{gru10}. In the present work, we
used a set of exposures taken between May 4$^{\rm th}$ and May 9$^{\rm th}$
2009.

Finally, we have included $\tau$ Sco (HD~149438) in our sample. Although not an O--type star per se, $\tau$ Sco is an early B dwarf (spectral type B0.2V) just at the limit of the O star mass range. Since a magnetic field of the order 300 G has been clearly detected on its surface \citep{donati06tausco}, it was justified to add it to our otherwise relatively limited sample. The geometry of the magnetic field is quite complex, with energy in multipoles up to order five. The origin of the field is debated, although a fossil origin is preferred. We have used a selection of spectra taken between 2005 and 2008 to conduct our analysis. 

To maximize the signal--to--noise ratios, we used the average of the
observations presented in Table\ \ref{tabA1} for all stars. This
resulted in spectra with SNR above 1000.

%%%%%%%%%%%%%%%%%%%%%%%%%%%%%%%%%%%%%%%%%%%%%%%%%%%%%%%%%%%%%%%%%%%%%%%%%%%%%%%%%%%%%%%%%%%%%%%%%%%%%%%%%%%%%%%%%%%%%%%%%%%%%%%
%%%%%%%%%%%%%%%%%%%%%%%%%%%%%%%%%%%%%%%%%%%%%%%%%%%%%%%%%%%%%%%%%%%%%%%%%%%%%%%%%%%%%%%%%%%%%%%%%%%%%%%%%%%%%%%%%%%%%%%%%%%%%%%
\section{Spectral analysis}
\label{s_mod}

%%%%%%%%%%%%%%%%%%%%%%%%%%%%%%%%%%%%%%%%%%%%%%%%%%%%%%%%
\subsection{Model Atmospheres}

The spectroscopic analysis was performed with the atmosphere code
CMFGEN \citep[a complete description of the code is provided
  by][]{hm98}. The models include stellar winds and are computed in
spherical geometry. The code solves simultaneously the radiative
transfer equation (in the co-moving frame) and the rate equations,
following an iterative scheme. The computations provide non-LTE
occupation numbers. The temperature structure is given by the
radiative equilibrium equation and the density structure by the mass
conservation equation.

Line-blanketing is included in CMFGEN through a super-level formalism
in order to reduce the computational cost. Our models include H, He,
C, N, O, Si, Ne, Mg, Si, S, Fe and Ni.  For each species, the solar
abundances from \cite{gre07} were taken as reference.

The wind structure is described by a mass loss rate, a terminal
velocity and a $\beta$--velocity law \citep{pauldrach86,kud89}.  The
subsonic part of the atmosphere is described by a pseudo-hydrostatic
structure, computed with the TLUSTY code \citep{lh03} with a
micro-turbulent velocity $\xi_{t}$ = 10 km s$^{-1}$. The velocity
structure of the atmosphere is then fully defined by connecting the
subsonic and the supersonic components. We have used clumped models, with a volume filling factor of 0.1 (0.01 for $\zeta$ Ori -- see \citet{bouret08} and HD108 -- see \citet{naze08b}). The other wind parameters have been adopted from the literature \citep{donati06tausco,bouret08,naze08b,grunhut09,wade191612,wade11}. 

Once a model atmosphere is converged, the final synthetic spectrum is
obtained from a formal solution of the radiative transfer equation.  A
depth-dependent micro-turbulent velocity $\xi_{t}$(r), equal to 5--10
km s$^{-1}$ at the base of the wind ($\xi_{0}$) and to about 10 $\%$
of the terminal velocity far from the photosphere ($\xi_{\infty}$), is
allowed. The value $\xi_{t}$ used for the models has very little
effect on the atmosphere structure, and consequently on the emergent
spectrum. On the other hand, $\xi_{0}$ may modify the shape and strength
of the photospheric lines. We found that a value of 10 \kms\ gave
satisfactory fits for most stars, except HD~57682 (the most
sharp--lined star of the sample) for which a value of 5~\kms\ was
better.

%%%%%%%%%%%%%%%%%%%%%%%%%%%%%%%%%%%%%%%%%%%%%%%%%%%%%%%%
\subsection{Stellar parameters}

\begin{figure*}[!ht]
\centering
\includegraphics[width=18cm]{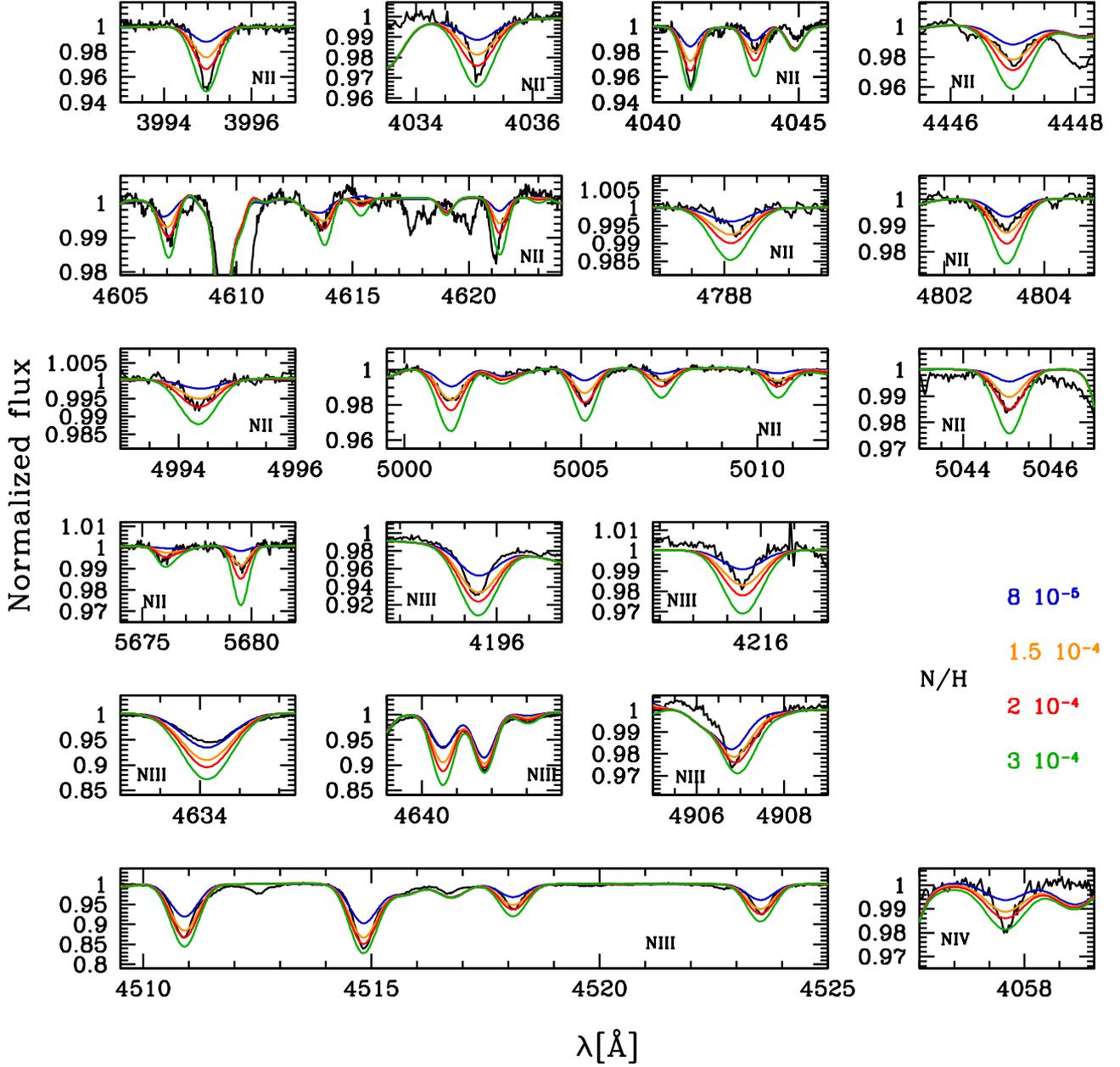}
\caption{Example of fit of the nitrogen lines in HD~57682. The observed spectrum is the black solid line. The smooth colored lines are models with the best fit parameters and different values of $N/H$. The $\chi^{2}$ function is evaluated using these models to constrain the value of surface nitrogen content. In the present case, $N/H = 1.3 \times 10^{-4}$ is the preferred value (see Fig.\ \ref{fig2}, red curve). More models with additional $N/H$ values have been calculated but only four are shown for clarity.}
\label{fig_nfit}
\end{figure*}

The stellar parameters of our best models are gathered in Table~\ref{tab_params}.
They correspond to the set of parameters that best fit the observed
spectrum of the respective sample star. The uncertainties
are indicated for each parameter. We will detail below how we
determined the uncertainties on the nitrogen abundance. The
errors on effective temperature and surface gravity depend on
the quality of the fits we obtain, and are used to compute the
errors on luminosity, radius and mass. The main source of
uncertainty on the luminosity is the distance.

We derived the effective temperature by the ionization balance
method. In practice, the relative strength of \ion{He}{i} and
\ion{He}{ii} lines was used. We relied preferentially on
\ion{He}{i}~$\lambda$4471 and \ion{He}{ii}~$\lambda$4542.  Additional
diagnostics were: \ion{He}{i}~$\lambda$4026,
\ion{He}{i}~$\lambda$4388, \ion{He}{i}~$\lambda$4712,
\ion{He}{i}~$\lambda$4920, \ion{He}{i}~$\lambda$5876,
\ion{He}{ii}~$\lambda$4200, \ion{He}{ii}~$\lambda$5412. In the Of?p
stars, the \ion{He}{i} lines are partly contaminated by some
additional emission the origin of which is unclear. The contamination
is the largest in the ``active'' states, i.e. when the
\ion{C}{III}~$\lambda$ 4650 lines have their maximum emission
strength. Consequently, the temperature determination is difficult in
those states and this explains why we rely on the quiet states
whenever possible.

The Balmer line wings are the best surface gravity diagnostics
available in optical spectra. We relied mainly on H$\beta$, H$\gamma$
and H$\delta$ to constrain \logg. In the Of?p stars, the Balmer lines
suffer from the same contamination as the \ion{He}{i} lines in the
active state. Fortunately, the wings, especially the blue one, are
relatively unaffected by this contamination, allowing a safe gravity
determination. Note that H$\delta$ is blended with NIII $\lambda$4097,
rendering the use of this line less suitable. It is used as a
secondary indicator.

The luminosities 
are computed in two ways. If calibrated UV flux and optical--
infrared photometry is available, the spectral energy
distribution is fitted. The Galactic extinction laws of \citet{seaton79}
and \citet{howarth83} are used to derive the extinction. The
luminosity is adjusted to match the SED, for a given distance. If
only optical photometry is available, we derived the luminosity
from the $V$ band magnitude, the effective temperature, the
bolometric correction \citep[from][]{mp06}, the extinction
(calculated from B-V) and the distance. Once known, the
luminosities directly provide the radii.

The distance of the Orion stars is set to $d = 414 \pm 50$ pc
\citep{menten07}. The distance to the Of?p stars is determined by
their supposed membership to OB associations, and was taken from
\citet{humphreys78}.  HD~108 belongs to the Cas~OB5 association. We
kept the luminosity derived by \citet{martins10} assuming a distance
of 2.51$\pm$0.10 kpc and fitting the spectral energy
distribution. HD~148937 is a member the Ara~OB1 association, at a
distance of 1.38$\pm$0.12 kpc.  HD~191612 belongs to the Cyg~OB3
association, at a distance of 2.29$\pm$0.12 kpc. $\tau$ Sco
  has a reliable Hippacros parallax leading to a distance of 145$\pm$11
  pc.

\citet{grunhut09} assumed the luminosity of HD~57682 to be
typical of a late O--type star and thus adopted the calibrated
value of \citet{msh05}. From this
luminosity and the fit to the IUE spectrum, they derived a distance
of 1.3 kpc. We adopted their luminosity for our own models

The projected rotational velocities and macroturbulent velocities have been determined from the fit of \ion{He}{i} 4713. A first guess of \vsini\ was obtained from the line Fourier transform. It usually led to an upper limit, no zero being observed above the noise level. We then used various combinations (\vsini\ / \vmac) to convolve synthetic spectra and selected the values best representing the observed line profile \citep[see also][]{howarth07}. We used a Gaussian profile to take macroturbulent velocities into account. 

\begin{table*}[t]
\centering
\caption{Stellar parameters derived from quantitative analysis with atmosphere models.}
\begin{tabular}{llllllll}
\hline
\hline
Star                  & HD191612          & $\Theta^{1}$ Ori C & $\zeta$ Ori A  & HD57682           & HD108              & HD148937        &  $\tau$ Sco \\
\hline
Spectral type         & O6--8f?p          & O7V               & O9.7Ib           & O9IV             & O4--8f?p          & O5.5--6f?p      & B0.2V \\
$T_{\rm eff}$ (K)          & 36000$\pm$2000    &  38000$\pm$2000   & 29500$\pm$1000   & 34500$\pm$1000   & 35000$\pm$2000    & 40000$\pm$2000  & 31000$\pm$1000 \\
log $g$               & 3.75$\pm$0.10     &   4.20$\pm$0.20   & 3.25$\pm$0.10    & 4.00$\pm$0.20    & 3.50$\pm$0.10     & 4.00$\pm$0.10   & 4.10$\pm$0.10 \\
%\hline
log ($L$ / L$_{\sun}$)  & 5.45$\pm$0.15      &  4.80$\pm$0.13    & 5.64$\pm$0.15   & 4.79$\pm$0.25    & 5.70$\pm$0.10      & 5.80$\pm$0.15  & 4.37$\pm$0.08 \\
$R_{*}$ (R$_{\sun}$)   & 12.3$\pm$1.9       &  5.8$\pm$1.2        & 25.4$\pm$1.7    & 7.0$\pm$0.4     & 19.4$\pm$2.2       & 16.6$\pm$1.7  & 5.3$\pm$0.6 \\
$M_{\rm ev}$ (M$_{\sun}$)  & 37.7$^{+3.7}_{6.2}$  & 23.8$^{+3.0}_{-2.9}$ & 42.8$^{+9.7}_{9.4}$& 22.0$^{+3.4}_{3.7}$ & 48.8$^{+5.4}_{-8.6}$ & 57.9$^{+7.0}_{-10.4}$ & 15.9$^{+0.6}_{-0.6}$ \\
$M_{\rm spec}$ (M$_{\sun}$) & 38.7$\pm$20.7     & 19.7$\pm$14.0     & 42.0$\pm$21.2     & 17.8$\pm$14.02    & 43.3$\pm$18.7     & 101.0$\pm$53.1  & 13.1$\pm$4.5 \\
$V$~sin$i$ (km s$^{-1}$)&  $<$ 1           & 24                &  100              & 15                & $<$ 1              & $<$ 50         &  6 \\
$v_{\rm mac}$ (km s$^{-1}$)& 40              & 30                &  $\sim$0          & 16                & 45               & 60               &  3 \\

$N/H$ (number)  & (2.7$\pm$1.5) 10$^{-4}$ & (5.1$\pm$4.5) 10$^{-5}$ & (3.3$\pm$1.0) 10$^{-5}$  & (1.3$\pm$0.3) 10$^{-4}$ & (5.9$\pm$4.2) 10$^{-4}$ & (3.0$\pm$1.2) 10$^{-4}$ & (1.4$\pm$0.2) 10$^{-4}$ \\
12+log($N/H$)        &  8.43$\pm$0.24       & 7.82$\pm$0.29     & 7.52$\pm$0.14     & 8.11$\pm$0.10    & 8.77$\pm$0.31     & 8.48$\pm$0.17 & 8.15$\pm$0.06 \\
\hline
\hline
\end{tabular}
\tablefoot{The evolutionary masses are determined using the tracks of \citet{brott11a} with an initial rotational velocity of 300 \kms.}
\label{tab_params}
\end{table*}

%%%%%%%%%%%%%%%%%%%%%%%%%%%%%%%%%%%%%%%%%%%%%%%%%%%%%%%%
\subsection{Surface nitrogen abundance}

The main purpose of the present work is to derive surface nitrogen
abundances as accurately as possible, in order to compare
them to the predictions from stellar evolution models. 

This task is challenging considering that the optical spectra
of O stars exhibit a small number of nitrogen lines, and that
most of these lines are relatively weak. In practice, we have used all nitrogen lines that our models were able to reproduce. Since we have a relatively heteregeneous sample of stars in terms of effective temperature, the number of lines used is different for each star. The only common lines are the \ion{N}{iii} 4510~-~4525 features. These lines are relatively free of contamination and do not suffer from severe blending. Their formation do not seem to be affected by line--blanketing effects. Their lower levels are connected to the \ion{N}{iii} ground level by transitions at 434\AA, a region relatively free of metallic lines. Hence, subtle radiative transfer effects such as those reported by \citet{paco06} are not expected. For HD~57682 and $\tau$~Sco we took advantage of the relatively large number of \ion{N}{ii} lines present in the observed spectra. For the Of?p stars, we only relied on \ion{N}{iii} features since the stars are too hot to display \ion{N}{ii} lines and \ion{N}{iv}~4058 was clearly not reproduced. This inconsistency is probably not related to an ionization issue since for stars with lower \teff\ the \ion{N}{iii}/\ion{N}{ii} balance matches the \ion{He}{ii}/\ion{He}{i} balances. \citet{rgonz11b} reported that in some of their models, they could not correctly reproduce this feature (as well as other \ion{N}{iv} lines). Part of the problem might be due to the sensitivity of these lines to wind properties \citep{rgonz11b}. We did not perform an investigation of the wind properties of these objects, and thus we refrained from using this line when clear inconsistencies were observed. In addition, since magnetic field severely complicates the determination of wind properties (due to geometry, clumping), it is very likely that \ion{N}{iv}~4058 is not perfectly predicted by our spherical models. Similar considerations can be applied to the strong \ion{N}{iii} 4634~-~4643 lines. \citet{rgonz11a} showed that the onset of the wind was critical in the formation of this triplet. We note in addition that blends with \ion{O}{ii} lines exist at 4643 \AA. We thus used these lines only when pure absorption was observed. Besides, the Of?p stars are known to display narrow \ion{N}{iii} 4634~-~4643 emission lines of unknown origin \citep[e.g.][]{naze08b}. In the case of $\Theta^{1}$ Ori C, we could reliably use only the \ion{N}{iii} 4510~-~4525 and \ion{N}{iii}~4095 features (the latter being located in the wing of H$\delta$ and thus subject to more uncertainties). All the other \ion{N}{iii} features were too weak.

\begin{figure}[h]
\centering
\includegraphics[width=88mm]{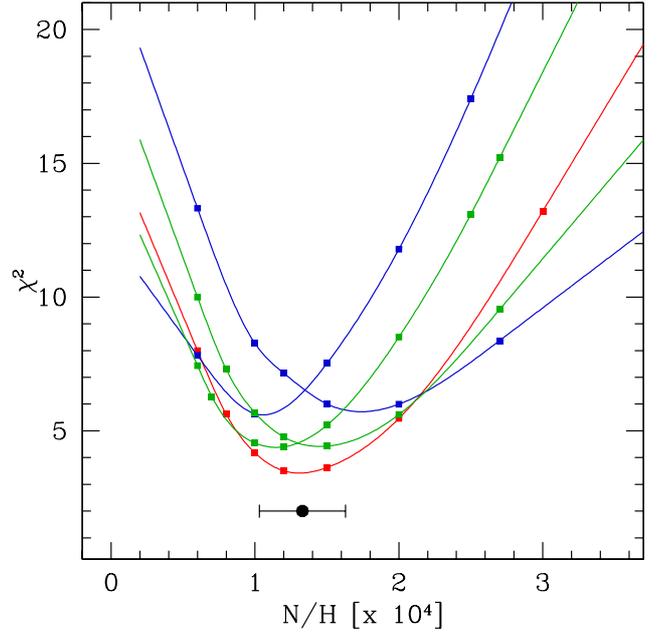}
\caption{Chi squares computed for the quiet state of HD~57682 as a function of the nitrogen abundance. Each color represents a
		parameter set: red for the best fit model (parameters given in Tab. \ref{tab_params}), green for variations of the surface gravity ($\pm$ 0.1 dex with respect to the best fit model) and finally,
		blue for variations of the effective temperature ($\pm$~1000 K with respect to the best fit model). The symbols correspond to the models computed.
		The different curves are polynomial fits to the respective set of symbols (red triangles, blue squares,$\dots$). The black dot with the error bars shows the position of the final $N/H$ value and its uncertainty.}
\label{fig2}
\end{figure}

Special care was taken to estimate the error bars on the derived surface nitrogen abundances. To do so, we proceeded as follows: for the best fit model (best values of \teff\ and \logg) we ran models at different nitrogen abundances. We then computed two new series of models with \teff\ increased (respectively decreased) by the uncertainty on \teff. We proceeded similarly for two series of models with the best estimate of \teff, but \logg\ increased (decreased) by the uncertainty on \logg. We ended up with  five series of models with 5--8 values of $N/H$ for each series. Fig.\ \ref{fig_nfit} shows an illustration of the model fits to the observations in the case of HD~57682. 

We then estimated the preferred $N/H$ content for each serie. For that, we calculated for the entire set of lines

\begin{equation}
 \chi^{2}= \frac{1}{N-1}\sum_{i=1}^{N} \frac{(M_{i}-S_{i})^{2}}{\sigma_{i}^{2}}
\end{equation}

\noindent where $M_{i}$ is the model spectrum, $S_{i}$ the observed spectrum, $\sigma_{i}$ the uncertainty on the observed spectrum and $N$ the number of wavelength points over which the function is estimated. We determined the value of $N/H$ that  minimizes the chi-square for each set of parameters (\teff, \logg). An illustration of $\chi^2$ versus $N/H$ for the various models computed for HD~57682 is given in Fig.~\ref{fig2}. 

To obtain the final value of $N/H$, we determined the best fit value for each combination (\teff, \logg) and computed the average and one $\sigma$ dispersion of this set of values. The final values are given in Table \ref{tab_params}.

To further check the robustness of our determinations, we ran a few additional tests. In particular, we tested the effect of a change of the microturbulent velocity in the atmosphere model computation (not just the output synthetic spectrum). Increasing this velocity from 5 to 10 \kms\ translates to a decrease of the derived abundance by 15--20\%. Although not negligible, this is within the uncertainties of our determinations. 

Abundance determination have been performed by \citet{morel11} for two our our sample stars: HD~57682 and $\tau$~Sco. For the latter, our results are in very good agreement (12+log($N/H$)=8.15$\pm$0.06 versus 8.15$\pm$0.20). \citet{przy10} also found a value of 8.16$\pm$0.12, in good agreement with our findings. For HD~57682, we derive a nitrogen content much larger than \citet{morel11} (8.11$\pm$0.10 versus 7.52$\pm$0.25). We attribute partially this discrepancy to the use of Kurucz models in the Morel et al. study. They did not self-consistently derive the stellar parameters and the abundances, but adopted \teff\ and \logg\ from the study of \citet{grunhut09}. Another cause of the problem can be a different line list used to perform the analysis. However, we do not have access to the lines used by Morel et al. and cannot test this possibility.

\begin{table*}[t]
\centering
\caption{Lines used to perform Nitrogen abundance determination.}
\begin{tabular}{lccccccc}
\hline
\hline
                         &  HD~191612  &$\Theta^{1}$ Ori C & $\zeta$ Ori A  & HD57682  & HD108    & HD148937  & $\tau$ Sco\\
\hline
\ion{N}{ii}~3995         &   --       &  --              &  $\surd$        &  $\surd$ & --       & --        & $\surd$ \\
\ion{N}{iii}~4004        &   --       &  --              &  --             &  --      & --       & $\surd$   & $\surd$ \\
\ion{N}{ii}~4035         &   --       &  --              &  --             &  $\surd$ & --       & --        & $\surd$ \\
\ion{N}{ii}~4041--4043   &   --       &  --              &  $\surd$        &  $\surd$ & --       & --        & $\surd$ \\
\ion{N}{iv}~4058         &   --       &  --              &  --             &  $\surd$ & --       & --        & $\surd$ \\
\ion{N}{iii}~4095        &   --       &  $\surd$         &  --             &  --      & --       & --        & --      \\
\ion{N}{iii}~4196        &   $\surd$  &  --              &  $\surd$        &  $\surd$ & $\surd$  & --        & $\surd$ \\
\ion{N}{iii}~4216        &   $\surd$  &  --              &  --             &  $\surd$ & $\surd$  & --        & $\surd$ \\
\ion{N}{ii}~4447         &   --       &  --              &  $\surd$        &  $\surd$ & --       & --        & $\surd$ \\
\ion{N}{iii}~4510--4525  &   $\surd$  &  $\surd$         &  $\surd$        &  $\surd$ & $\surd$  & $\surd$   & $\surd$ \\
\ion{N}{iii}~4537        &   $\surd$  &  --              &  --             &  --      & $\surd$  & $\surd$   & $\surd$ \\
\ion{N}{ii}~4602         &   --       &  --              &  --             &  --      & --       & --        & $\surd$ \\
\ion{N}{ii}~4607         &   --       &  --              &  --             &  $\surd$ & --       & --        & $\surd$ \\
\ion{N}{ii}~4613         &   --       &  --              &  --             &  $\surd$ & --       & --        & $\surd$ \\
\ion{N}{ii}~4621         &   --       &  --              &  --             &  $\surd$ & --       & --        & $\surd$ \\
\ion{N}{ii}~4634--4643   &   --       &  --              &  --             &  $\surd$ & --       & --        & $\surd$ \\
\ion{N}{ii}~4788         &   --       &  --              &  --             &  $\surd$ & --       & --        & $\surd$ \\
\ion{N}{ii}~4803         &   --       &  --              &  --             &  $\surd$ & --       & --        & $\surd$ \\
\ion{N}{iii}~4907        &   --       &  --              &  --             &  $\surd$ & --       & --        & -- \\
\ion{N}{ii}~4995         &   --       &  --              &  --             &  $\surd$ & --       & --        & $\surd$ \\
\ion{N}{ii}~5000--5011   &   --       &  --              &  $\surd$        &  $\surd$ & --       & --        & $\surd$ \\
\ion{N}{ii}~5026         &   --       &  --              &  --             &  --      & --       & --        & $\surd$ \\
\ion{N}{ii}~5045         &   --       &  --              &  --             &  $\surd$ & --       & --        & $\surd$ \\
\ion{N}{ii}~5676--5679   &   --       &  --              &  $\surd$        &  $\surd$ & --       & --        & $\surd$ \\
\hline
\end{tabular}
\label{tab2}
\end{table*}

%%%%%%%%%%%%%%%%%%%%%%%%%%%%%%%%%%%%%%%%%%%%%%%%%%%%%%%%%%%%%%%%%%%%%%%%%%%%%%%%%%%%%%%%%%%%%%%%%%%%%%%%%%%%%%%%%%%%%%%%%%%%%%%
%%%%%%%%%%%%%%%%%%%%%%%%%%%%%%%%%%%%%%%%%%%%%%%%%%%%%%%%%%%%%%%%%%%%%%%%%%%%%%%%%%%%%%%%%%%%%%%%%%%%%%%%%%%%%%%%%%%%%%%%%%%%%%%
\section{Stellar magnetism and nitrogen enrichment}
\label{s_mag}
	
%%%%%%%%%%%%%%%%%%%%%%%%%%%%%%%%%%%%%%%%%%%%%%%%%%%%%%%%%%%%%%%%%%%%%%%%%%%%%%%%%%%%%%%%%%%%%%%%%%%%%%%%%%%%%%%%%%%%%%%%%%%%
\subsection{Comparison to predictions of rotating models without magnetic field}

\begin{figure}[t]
\centering
\includegraphics[width=9cm]{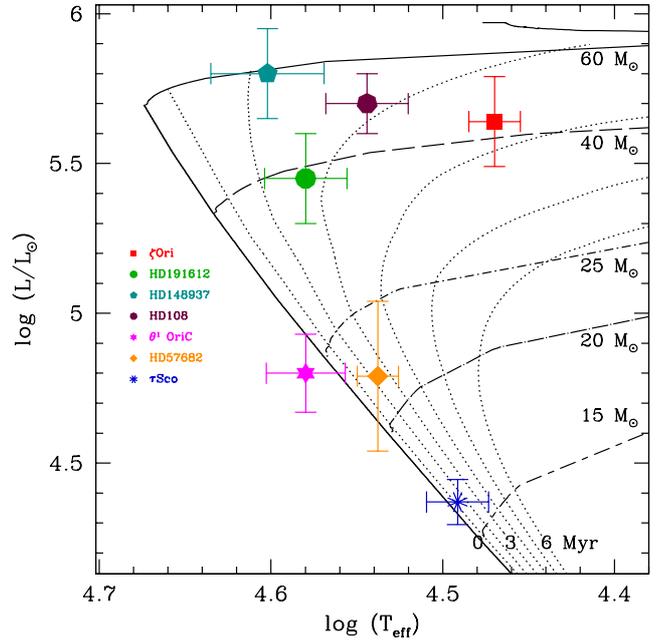}
\caption{Hertzsprung-Russel diagram for the sample magnetic OB stars. The isochrones and the evolutionary
		tracks are from \citet{brott11a} and were computed with an initial rotational velocity of 300 km s$^{-1}$.}
\label{hr300}
\end{figure}

\begin{figure}[t]
\centering
\includegraphics[width=9cm]{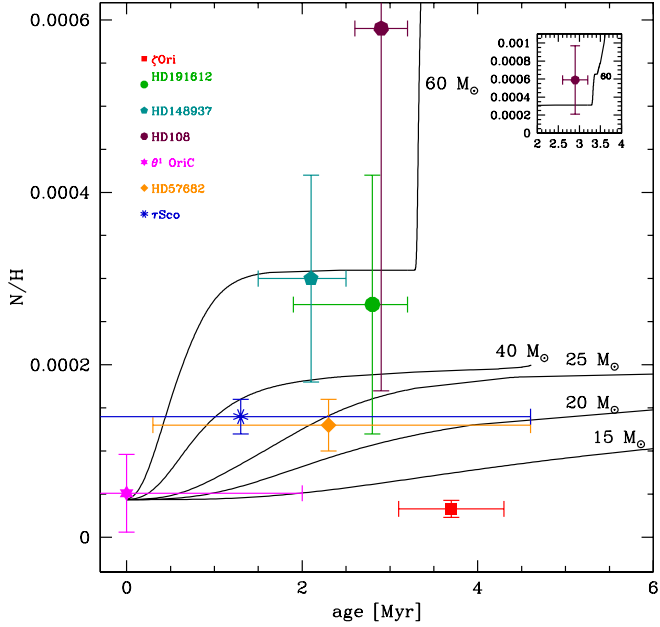}
\caption{$N/H$ as a function of age. Black solid lines are the predictions of evolutionary models including rotational mixing with an initial rotation rate of 300\kms. Models are from \citet{brott11a}. The symbols are the magnetic O stars studied in the present paper.}
\label{n_age}
\end{figure}

In Fig.\ \ref{hr300} we show the Hertzsprung--Russel (HR) diagram with the positions of the magnetic OB stars analyzed in the present study. We have chosen the evolutionary tracks and isochrones of \citet{brott11a}. Other evolutionary tracks exist in the literature, but the Brott et al. tracks have been computed for a large set of initial rotational velocities, allowing an investigation of the effects of rotation on the results. In addition, they have the same initial metal content as that used in our atmosphere models, rendering possible comparison between our results and the predictions of evolutionary calculations. Compared to the recent Geneva tracks \citep{ekstroem11} they predict a stronger nitrogen enrichment at a given age (for moderate and zero rotation) during the early evolution representing our sample stars. Consequently, our conclusions regarding a possible extra enrichment of magnetic stars would be re-enforced if we were using the Geneva tracks (see below). In Fig.\ \ref{hr300}, we have selected the evolutionary tracks with an initial rotational velocity of 300 \kms. This value is typical since it is supposed to lead to an average rotation on the main sequence similar to that observed \citep{mm00}. 

From the HR diagram, one can estimate the evolutionary mass of each individual star. This is done by simple interpolation between the available tracks. The results are given in Table~\ref{tab_params}. The uncertainties reflect the errors on the effective temperature and luminosity. The age of the stars can also be tentatively derived. The ages are just a way of evaluating the degree of evolution of a given star. They are directly used to build Fig.~\ref{n_age} where we show the surface nitrogen content as a function of age. This plot, together with the HR diagram, can tell us if a star is evolving according to the predictions of standard evolutionary models, or if significant deviations are detected.

Let us take HD~148937 as an example. From the HR diagram, its evolutionary mass should be 57.9$^{+7.0}_{-10.4}$ \msun. The star is 2.8$^{+0.4}_{-0.9}$ Myr old. Looking at Fig.\ \ref{n_age}, we see that for this range of ages, the surface nitrogen content of HD~148937 is consistent with stars of initial mass 60 \msun. This is thus consistent with the estimated initial mass of HD~148937, from which one can conclude that this star does seem to have normal surface abundance for its evolutionary status. A similar conclusion can be drawn for HD~191612 and HD~108, although for the latter a hint of extra enrichment is not excluded. 

$\tau$ Sco has an initial mass of about 16 \msun. Compared to the 15 \msun\ evolutionary tracks, its nitrogen content is rather high. Only for an effective temperature at the low limit of the derived range (and thus a relatively old age) can the N content be barely reconciled with the predictions. HD~57682 also has a nitrogen content 2 to 3 times the initial value. This value is marginally compatible with a 22 \msun\ star provided its age is on the upper side of the range of possible values. Hence, $\tau$ Sco and HD~57682 quite possibly show some extra enrichment. $\Theta^{1}$Ori C is very similar in mass but it is too young to have produced a significant amount of nitrogen and thus the low $N/H$ we derive is not unexpected.

The last star of our sample, $\zeta$ Ori A, is the oldest and most evolved of the magnetic O stars. At an age of 3.7$\pm$0.6 Myr and with an initial mass of 42.8$^{+9.7}_{-9.4}$, its surface $N/H$ should reach at least $2.0 \times 10^{-4}$. We derive a value about 4 times lower ($4.4\pm1.0 \times 10^{-5}$). As already noted by \citet{bouret08}, $\zeta$ Ori A is thus particularly chemically unevolved for its age.
 
\begin{figure*}[t]
     \centering
     \subfigure[v=100 \kms]{
%          \label{fig1}
          \includegraphics[width=.45\textwidth]{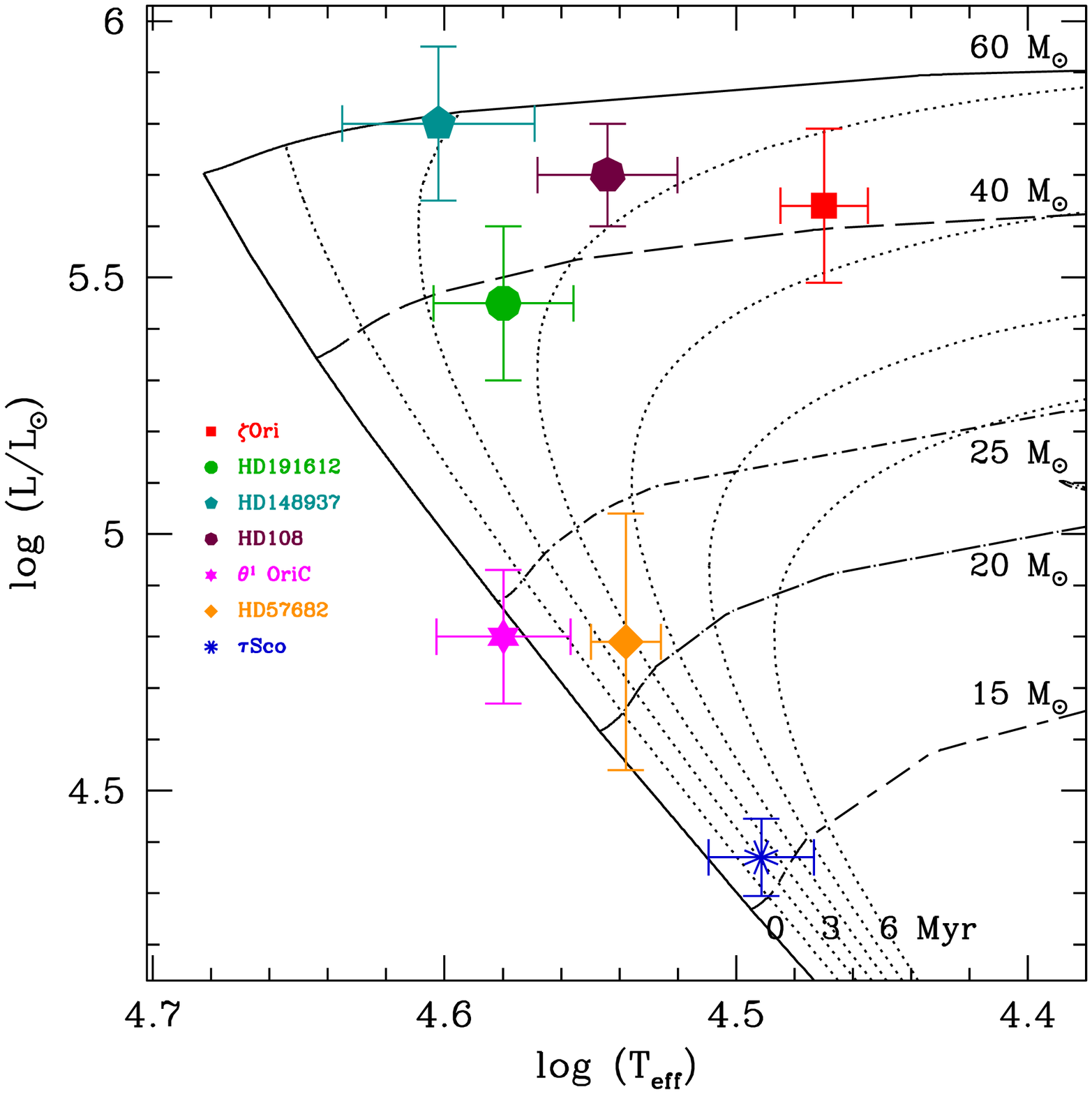}}
     \hspace{0.2cm}
     \subfigure[v=100 \kms]{
%          \label{fig2}
          \includegraphics[width=.45\textwidth]{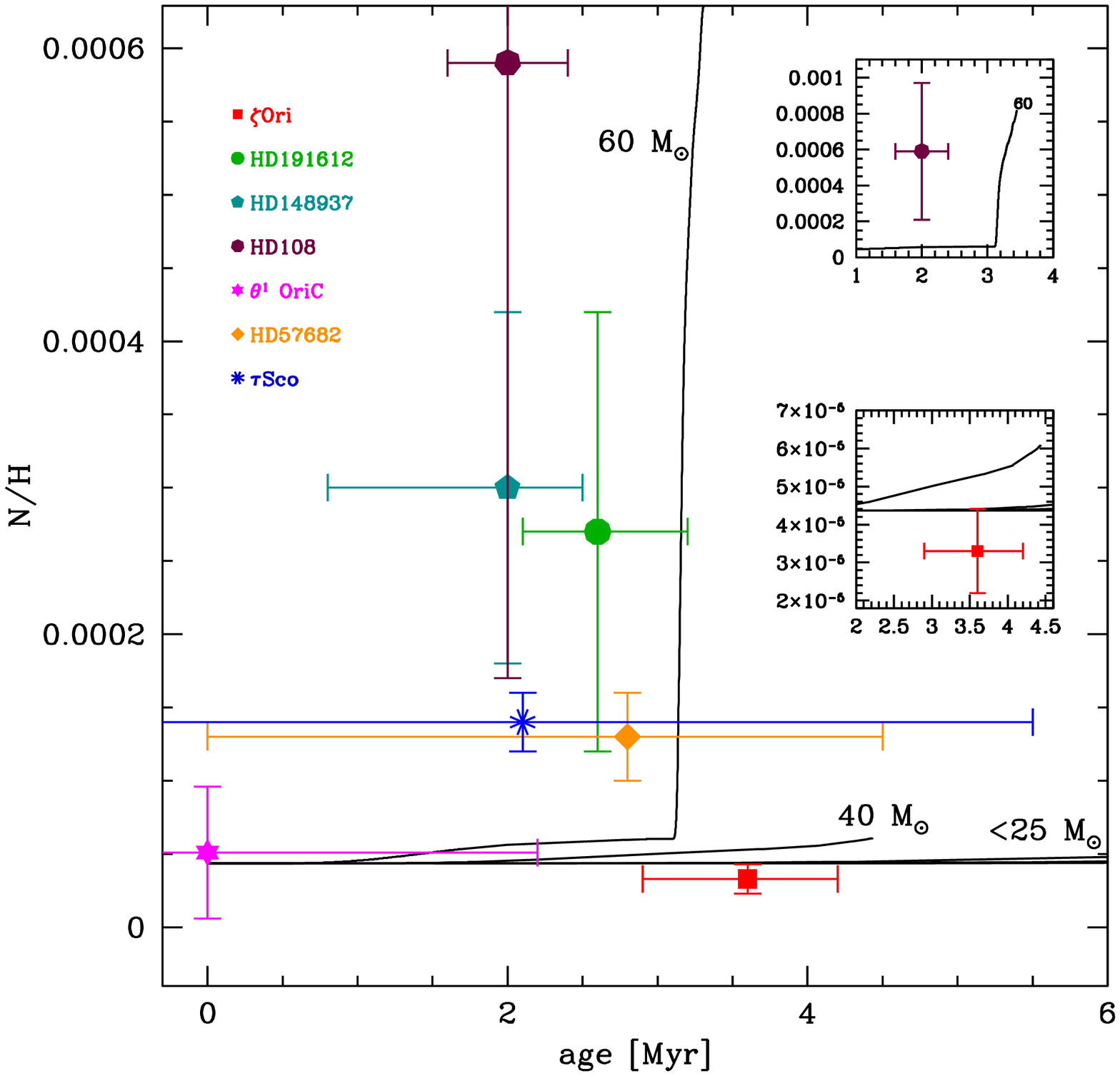}}
     \caption{Same as Figs.\ \ref{hr300} and \ref{n_age} but for initial rotational velocities of 100 \kms. The initial masses derived from the evolutionary tracks are gathered in Table \ref{tab_mev}.}
     \label{eff_rot}
\end{figure*}

So far, we have compared the properties of the magnetic O stars to evolutionary models with an initial rotational velocity of 300 \kms. This value leads to an average velocity on the main sequence consistent with observations of normal O stars. However, one might wonder if this value is appropriate for magnetic stars. The presence of a magnetic field is known to reduce the rotation rate as illustrated by \citet{townsend10} in the case of $\sigma$ Ori E. \citet{donati06} and \citet{martins10} have also noted the extremely low rotation of HD~191612 and HD~108. The question is: were these stars slow rotators already on the zero age main sequence? If so, the comparison with low \vsini\ evolutionary tracks is necessary. Table\ \ref{tab_tspin} presents the spin down timescales due to magnetic braking calculated according to \citet{ud09}. We have also listed the ages estimated from Figs.\ \ref{hr300} and \ref{eff_rot}. As noted by \citet{donati06} and \citet{martins10}, HD~191612 and HD~108 have spin down timescales marginally compatible with their age, implying that they might have experienced a reduction of their rotation rate by a factor 2--3. However, given their very low present--day \vsini, those stars probably had initial projected rotational velocities not higher than a few tens of \kms. All the other magnetic O stars have spin down timescales longer than their age. Given their rather low \vsini\ (with the exception of $\zeta$ Ori~A), these stars were also most likely slow rotators on the zero age main sequence. The youngest star of our sample ($\Theta^{1}$Ori C) confirms this trend: it is located close to the ZAMS and features a low \vsini\ (by O star standards).

In view of these comments, we show in Fig.\ \ref{eff_rot} the HR and $N/H$--age diagrams with evolutionary tracks computed for an initial rotational velocity of 100 \kms. Using these plots, we can see how the effects of rotation affect the previous results. The evolutionary masses derived from these new tracks are given in Table \ref{tab_mev}. The three Of?p stars (HD~108, HD~148937 and HD~191612) all appear to be much more N--rich than predicted by the 60 \msun\ evolutionary track, and thus more enriched than expected for their mass. $\tau$ Sco is now very clearly enriched compared to the 15 \msun\ tracks which barely departs from the initial N content in the first 6 Myr. For HD~57682, an extra enrichment is also clearly present. For $\Theta^{1}$Ori C, again the young age prevents any firm conclusion other than that no dramatic nitrogen enrichment is observed. For $\zeta$ Ori A the surface chemical pattern is lower than predicted by the evolutionary tracks although marginal agreement exists if extreme stellar parameters are used.

\begin{table*}[t]
\centering
\caption{Magnetic spin--down timescales and ages.}
\begin{tabular}{lcccccccc}
\hline
\hline
Star                  & $Bp$      & $M$      & $R$      &  \vinf\  &  \mdot\          &  \tspin    & age(300)   &  age(100) \\
                      & [kG]    & \msun\ & \rsun\ &  \kms\   &  [$10^{-7}$ \myr] & [Myr]      & [Myr]      & [Myr]   \\
\hline
HD~191612             & 2.5      & 37.7   & 12.3   &  2400    &  0.3        & 3.8            & 2.8$^{+0.4}_{-0.9}$  & 2.6$^{+0.6}_{-0.5}$ \\       
HD~148937             & 1.0      & 57.9   & 16.6   &  2600    &  0.3        & 11.3           & 2.1$^{+0.4}_{-0.6}$  & 2.0$^{+1.2}_{-0.5}$ \\ 
HD~108                & 0.5--2   & 48.8   & 19.4   &  2000    & 1.0         & 2.0--7.8       & 2.9$^{+0.3}_{-0.3}$  & 2.0$^{+0.4}_{-0.4}$ \\ 
$\Theta^{1}$ Ori C     & 1.1      & 23.8   & 6.5    &  2500    & 3.0         & 3.24           & 0.0$^{+2.0}_{-1.0}$  & 0.0$^{+2.2}_{-1.0}$ \\ 
$\zeta$ Ori A         & 0.05--0.1 & 42.8  & 25.4   &  2100    & 17.0        & 6.5--13.0      & 3.6$^{+0.6}_{-0.6}$  & 3.6$^{+0.7}_{-0.6}$\\ 
HD~57682              & 1.7      & 22.0   & 7.0    &  1200    & 0.014       & 18.8           & 2.3$^{+2.0}_{-2.3}$  & 2.8$^{+1.7}_{-2.8}$  \\
$\tau$ Sco            & 0.5      & 15.9   & 5.3    &  2000    & 0.3         & 17.0           & 1.3$^{+3.3}_{-2.3}$  & 2.1$^{+3.4}_{-2.6}$  \\
\hline
\hline
\end{tabular}
\tablefoot{The magnetic spin--down timescale is calculated according to \citet{ud09}. $Bp$ is the polar magnetic field. The stellar and wind parameters are used to calculate \tspin. We have assumed a value of 0.1 for the moment of inertia constant $k$ \citep[see Eq.\ 25 of][]{ud09}. The entry 'age(300)' is the age of the star estimated from the HR diagram built with evolutionary tracks computed for an initial rotational velocity of 300 \kms. Similarly, 'age(100)' is the age for \vsini\ = 100 \kms.}
\label{tab_tspin}
\end{table*}

\begin{table}[t]
\centering
\caption{Evolutionary masses derived for different initial rotational velocities.}
\begin{tabular}{llllllll}
\hline
\hline
Star                  & $M_{\rm ev}$ (300)      & $M_{\rm ev}$ (100)   \\  
\hline
\noalign{\smallskip}
HD~191612             & 37.7$^{+3.7}_{6.2}$   & 37.5$^{+6.1}_{6.2}$  \\
HD~148937             & 57.9$^{+7.0}_{-10.4}$ & 58.7$^{+9.2}_{-12.3}$ \\
HD~108                & 48.8$^{+5.4}_{-8.6}$  & 47.1$^{+7.6}_{-6.6}$  \\
$\Theta^{1}$ Ori C     & 23.8$^{+3.0}_{-2.9}$  & 23.6$^{+2.9}_{-3.0}$  \\
$\zeta$ Ori A         & 42.8$^{+9.7}_{9.4}$   & 43.0$^{+10.1}_{8.4}$  \\
HD~57682              & 22.0$^{+3.4}_{3.7}$   & 21.5$^{+3.6}_{5.3}$   \\
$\tau$ Sco            & 15.9$^{+0.6}_{0.6}$   & 15.7$^{+0.9}_{1.7}$   \\
\noalign{\smallskip}
\hline
\hline
\end{tabular}
\tablefoot{The models used are from \citet{brott11a}.}
\label{tab_mev}
\end{table}

Given the above discussion, the following conclusions can be drawn:

\begin{itemize}

\item $\zeta$ Ori A: For moderate rotation rate, the low N content cannot be explained by standard mixing. Some extra process leading to the quenching of rotational mixing is necessary to explain the peculiar $N/H$ ratio. The conclusion remains valid for low rotation rates, although the agreement with the expected $N/H$ ratio is marginal given the uncertainties on the age and surface nitrogen content.

\item HD~108, HD~148937, HD~191612, $\tau$ Sco and HD~57682: under the reasonable assumption that these stars were slow rotators on the zero main sequence (i.e. initial rotation velocity of a few tens of \kms), all stars show a higher degree of nitrogen enrichment than non--magnetic evolutionary models predict. In the less likely case where the initial rotational velocity was normal (i.e. around 300 \kms), HD~108, HD~57682, $\tau$ Sco and HD~191612 show marginal evidence for extra enrichment, while HD~148937 appears normal.

\item $\Theta^{1}$Ori C: this star does not show peculiar nitrogen enrichment whatever the initial rotational speed. This is most likely because it is too young.

\end{itemize}

%%%%%%%%%%%%%%%%%%%%%%%%%%%%%%%%%%%%%%%%%%%%%%%%%%%%%%%%%%%%%%%%%%%%%%%%%%%%%%%%%%%%%%%%%%%%%%%%%%%%%%%%%%%%%%%%%%%%%%%%%%%%
\subsection{Comparison to non--magnetic stars}

\begin{figure}[t]
\centering
\includegraphics[width=9cm]{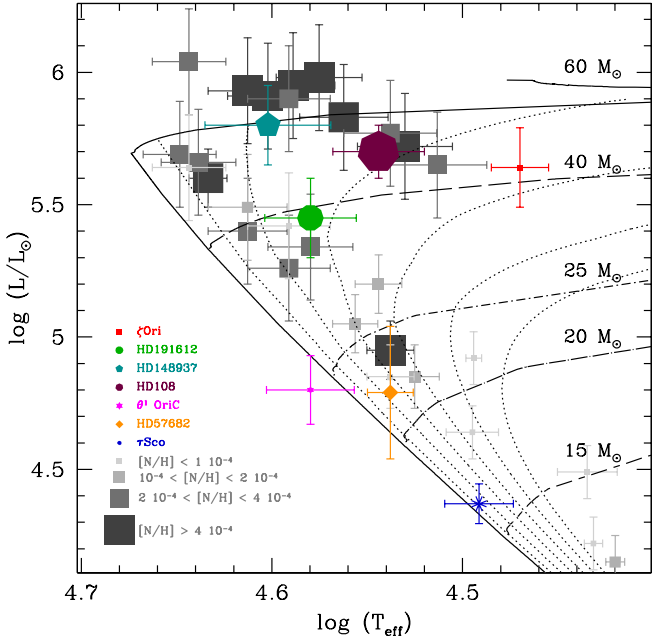}
\caption{Same as Fig.\ \ref{hr300} with comparison stars added (grey squares). The comparison stars are from \citet{jc05}, \citet{ngc2244}, Bouret et al.\ (2012, A\&A, in prep), Escolano et al.\ (2012, A\&A, in prep), \citet{hunter09} and \citet{przy10}. The size of the symbols (including the six magnetic O stars) is now proportional to the $N/H$ value. The evolutionary tracks are for $v = 300$ \kms.}
\label{hr_comp}
\end{figure}

To further investigate the surface chemical properties of the six magnetic O stars, we show in Fig.\ \ref{hr_comp} the same HR diagram as in Fig.\ \ref{hr300} to which we have added comparison O stars. The latter objects have been taken from \citet{jc05}, \citet{ngc2244}, Bouret et al.\ (2012, A\&A, in prep), Escolano et al.\ (2012, A\&A, in prep), \citet{przy10} and \citet{hunter09}. All O stars have been analyzed with CMFGEN in the same way as in the present study. In Fig.\ \ref{hr_comp} we encode the value of $N/H$ by the size of the symbol for each star (including the magnetic O stars). In the upper part of the HR diagram, HD~108 and HD~148937 fall in a densely populated group of O supergiants with $N/H$ larger than $2.0 \times 10^{-4}$. They thus appear as similarly enriched compared to these objects, and present no peculiarity in their chemical pattern provided they were initially rotating at about 300 \kms\ (but see below).

HD~191612 is located at a lower luminosity than the two other Of?p stars. In that region of the HR diagram, the surface nitrogen abundance of the few O stars present a large range of values, from less than $10^{-4}$ to more than $4.0 \times 10^{-4}$. In addition, the less evolved stars are more enriched than the more evolved objects. Given these puzzling results, the surface abundance of HD~191612 do not appear to deviate from the general trend, but the large range of $N/H$ values observed in that part of the diagram prevents any firm conclusion regarding its 'normality'.

The last star in the upper part of the HRD is $\zeta$ Ori A. Its surface nitrogen content strongly deviates from the values observed in neighboring stars. The O supergiants analyzed so far all show significant N enrichment ($N/H > 2.0 \times 10^{-4}$) while $\zeta$ Ori A is basically chemically unevolved. This confirms our previous conclusions that this star clearly shows peculiar chemical properties.

Going down in the HR diagram, we find $\Theta^{1}$Ori C and HD~57682 very close to each other. Only three O stars are located close enough to allow a quantitative comparison. These objects show a wide range of $N/H$ values, from less than 1.0 10$^{-4}$ to more than $4.0 \times 10^{-4}$. Both $\Theta^{1}$Ori C and HD~57682 show relatively modest enrichment. This is consistent with the two less enriched comparison stars. Hence, the two magnetic stars do not seem to stand out as peculiar objects. Finally, close to the location of $\tau$ Sco we have found very few comparison stars and so we cannot draw any relevant conclusion.  

At first sight, this comparison only exacerbates the peculiar properties of $\zeta$ Ori A. All the other stars appear similar to non--magnetic O stars. There is however an important property of the magnetic stars to consider: all (except $\zeta$ Ori A) are slow rotators. As explained above, there are indications that they might have been slow rotators already on the ZAMS. As such, they are different from the comparison stars which have diverse values of \vsini\ (32 stars with \vsini\ from roughly 0 to 330 \kms\ with an average of 116 \kms). Consequently, finding nitrogen enrichment in HD~108 and HD~148937 (the two objects for which there are sufficient comparison stars to draw firm conclusions) similar to that of comparison stars while they are rotating more slowly can be interpretated as the sign of an especially efficient mixing inside the star. Said differently, the (presumably) initially slowly rotating magnetic stars have the same surface N content as normal O stars. This is an indirect indication that the presence of a magnetic field in those objects tends to favor the transport of nitrogen from the core towards the surface.

Clearly, a larger sample of magnetic and comparison stars would be needed to strengthen these conclusions. But the above comparisons all indicate that there seems to be two groups of magnetic stars: on one side $\zeta$ Ori A which is clearly peculiar, with no nitrogen enrichment in spite of its evolved status and relatively normal rotational velocity; and the other OB stars, especially the Of?p objects, possibly displaying larger enrichment than non--magnetic stars provided they were indeed slow rotators on the ZAMS.

%%%%%%%%%%%%%%%%%%%%%%%%%%%%%%%%%%%%%%%%%%%%%%%%%%%%%%%%%%%%%%%%%%%%%%%%%%%%%%%%%%%%%%%%%%%%%%%%%%%%%%%%%%%%%%%%%%%%%%%%%%%%
\subsection{On the nature of magnetic field in O stars}

The origin of magnetic field in massive stars is a matter of debate. The two main possibilities are either a fossil field or a field generated by dynamo processes in the radiative region or the convective core. In the former case, the field might come from the parental molecular cloud, the convective phases during pre-main sequence evolution, of from binarity. Relatively simple geometries are expected in that case. For dynamo generated magnetic fields, several scenarios exists but the Tayler--Spruit dynamo has been the most tested \citep{spruit99,spruit02}. In this case, an initial poloidal field is transfered into a toroidal field by differential rotation. Later on, the Pitts--Tayler instability develops in the toroidal field and gives birth to a poloidal field. Consequently, the dynamo loop is closed. Controversy exists at present to know if this process is effectively at work. In a series of magneto--hydrodynamical (MHD) simulations, \citet{brait06} claim to have observed the onset of the Pitts--Tayler instability and the dynamo loop. But \citet{zahn07} conclude differently in their simulations: although the Pitts--Tayler instability seems present, it has no effect on the poloidal field which slowly decreases due to Ohmic decay. Hence the dynamo loop is not closed. A good summary of these processes is provided by \citet{maeder09} and \citet{walder11} \citep[see also][]{zahn11}.

Can we provide any constraint on the nature of O star magnetism? \citet{mm05} ran evolutionary calculations under the assumption that the Tayler--Spruit dynamo was working. In that case, they could show that one of the main effects was an increase of the mixing and of the efficiency of chemical element transport. The reason is the nearly solid body rotation achieved in the radiative zone in the presence of magnetic field. In non magnetic rotating models, chemical elements are mainly transported by shear turbulence caused by differential rotation. Meridional circulation -- created by thermal imbalance due to gravity darkening -- contributes little to the transport of elements, but is the main way of redistributing angular momentum. If the star rotates nearly solidly, shear turbulence is reduced, but meridional circulation is greatly enhanced so that the net effect on the transport of chemical species is a much stronger mixing. \citet{mm05} report an increase of the surface $N/H$ of a factor 2 to 5 larger than in non magnetic rotating models for a 15 \msun\ star of age 5 to 10 Myr. \citet{maeder09} note that these results depend on the details of the dynamo process at work and on the strength of the magnetic field. However, the trend seems to be an significant over--enrichment compared to standard rotating models. In their more recent investigation of the role of magnetic braking, \citet{meynet11} note that if the star is rigidly rotating, braking is so strong that it suppresses all surface chemical enrichment. Thus the exact role of magnetic field on surface abundances is not clear.

Our results are meant to provide observational constraints that could trigger further theoretical developments. At present, they can confront theoretical predictions as follows:

\begin{itemize}

\item dynamo fields: the evolutionary calculations focus on the Tayler--Spruit dynamo as explained above. One of the expectations of dynamo fields is that, similarly to the sun, they show complex morphologies and not simple dipolar structures. The constraints on the field geometry of magnetic O stars are still scarce, but $\zeta$~Ori~A stands out one more time \citep{bouret08} since the first observations indicates the possible presence of several magnetic spots. If a dynamo process is producing the magnetic field of that star, then our results indicate that no surface enrichment is observed. According to \citet{meynet11} -- \citep[see also][]{zahn11} -- this could be the sign of a nearly solid--body rotation and strong braking (see above). However, the projected rotational velocity of $\zeta$ Ori~A is not especially low (\vsini\ = 110 \kms). This is not consistent with a strong braking unless the star was initially rotating very fast. But in that case, would the high level of rotation be compatible with the absence of surface chemical enrichment? Clearly, dedicated simulations are necessary to see if a star with a dynamo field can still be rotating relatively normally and not have produced any surface nitrogen excess after several Myrs.

\item fossil fields: it is usually thought that fossil fields have simple dipolar geometries \citep{dl09}. Observations indicate that $\Theta^{1}$Ori C, HD~57682, HD~191612 and HD~148937 have dipolar fields. For HD~108, the phase coverage is too poor to provide any constraint. $\tau$ Sco features a more complex magnetic geometry, but \citet{donati06tausco} attributes this to the youth of the star. If a dipole should remain after a few Myrs in fossil field magnetic stars, higher orders are expected to remain in the early evolution, after the formation from the parental molecular cloud. \citet{donati06tausco} thus favoured a fossil origin for the magnetic field of $\tau$ Sco. 
According to Sect.\ \ref{s_mag}, all those objects were initially slow rotators on the ZAMS. Consequently, the observed surface abundances are higher than they would be if there was no magnetic field. In practice, the degree of enrichment corresponds to that observed in stars with initial \vsini\ of 300 \kms, while the initial rotational velocity of the magnetic stars was more likely of a few tens of \kms. Unfortunately, there are no simulations of the effect of fossil fields on the chemical appearance of massive stars to which we can compare these results. We can thus simply state that dipolar, probably fossil fields tend to lead to extra mixing in the interior of O stars. According to the results of \citet{meynet11}, this requires the presence of strong shear created by differential rotation. Magnetic O stars with fossil fields might thus have a steep velocity structure.

\end{itemize}

Clearly, the effects of magnetic field on the properties of O stars are still poorly understood. A larger sample of magnetic stars would be needed to see if the apparent separation into two groups (slowly rotating enriched stars / normally rotating non--enriched stars) is confirmed. Analysis of more comparison stars is required too. In parallel, new MHD simulations and evolutionary models, especially including fossil fields, are necessary to better understand the physics at work in the interior and atmosphere of magnetic O stars.

%%%%%%%%%%%%%%%%%%%%%%%%%%%%%%%%%%%%%%%%%%%%%%%%%%%%%%%%%%%%%%%%%%%%%%%%%%%%%%%%%%%%%%%%%%%%%%%%%%%%%%%%%%%%%%%%%%%%%%%%%%%%
%%%%%%%%%%%%%%%%%%%%%%%%%%%%%%%%%%%%%%%%%%%%%%%%%%%%%%%%%%%%%%%%%%%%%%%%%%%%%%%%%%%%%%%%%%%%%%%%%%%%%%%%%%%%%%%%%%%%%%%%%%%%
\section{Summary and conclusions}
\label{s_conc}

We have conducted a quantitative spectroscopic analysis of the six magnetic O stars known to date together with the early B star $\tau$ Sco. Spectra collected with ESPADONS at CFHT and NARVAL at TBL have been used. Atmosphere models computed with the code CMFGEN have been computed. We have determined the stellar parameters and the surface nitrogen abundance. A careful investigation of the uncertainties associated with $N/H$ has been pursued. 

Our results can be divided into two parts:

\begin{itemize}

\item[$\bullet$] The surface $N/H$ of $\zeta$ Ori A -- the star with the weakest magnetic field -- is lower than the predictions of evolutionary models. For that star the presence of the magnetic field probably quenches the transport of chemical elements. The conditions under which such a process can happen are not clear. Evolutionary models accounting for magnetic braking predict no chemical enrichment in solid body rotating stars. But their rotational velocities are also rather low, whereas  $\zeta$~Ori~A appears to have a rather standard \vsini. 

\item[$\bullet$] The other magnetic O stars, especially the three Of?p stars, display surface nitrogen content consistent with those of non-magnetic O stars with initial \vsini\ of 300 \kms. However, these magnetic O stars were most likely slow rotator on the zero age main sequence (with values of \vsini\ of a few tens of \kms\ at most). Consequently, they feature larger surface $N/H$ than they should if they did not host magnetic fields as also indicated by their nitrogen excess compared to rotating evolutionary models.

\end{itemize}

The presence of a magnetic field in O stars thus appears to produce two types of chemical peculiarities: either an extra enrichment, or the absence of enrichment. Confrontations with evolutionary predictions in the case of dynamo fields indicate that solid body rotation might explain the properties of the non--enriched star, but its relatively normal rotational velocity is a puzzle in that context. For the N--rich magnetic O stars, a fossil field is usually assumed. Our results should be a basis for comparison with future evolutionary calculations including fossil fields since such evolutionary models do not exist at present.

%%%%%%%%%%%%%%%%%%%%%%%%%%%%%%%%%%%%%%%%%%%%%%%%%%%%%%%%%%%%%%%%%%%%%%%%%%%%%%%%%%%%%%%%%%%%%%%%%%%%%%%%%%%%%%%%%%%%%%%%%%%%
%%%%%%%%%%%%%%%%%%%%%%%%%%%%%%%%%%%%%%%%%%%%%%%%%%%%%%%%%%%%%%%%%%%%%%%%%%%%%%%%%%%%%%%%%%%%%%%%%%%%%%%%%%%%%%%%%%%%%%%%%%%%
\begin{acknowledgements}
We thank the referee, Jo Puls, for valuable comments and for suggesting to add the star $\tau$ Sco to our sample. Many thanks to John Hillier for making his code CMFGEN available and for constant help with it. GAW acknowledges Discovery Grant support from the Natural Science and Engineering Research Council of Canada. 
\end{acknowledgements}

%%%%%%%%%%%%%%%%%%%%%%%%%%%%%%%%%%%%%%%%%%%%%%%%%%%%%%%%%%%%%%%%%%%%%%%%%%%%%%%%%%%%%%%%%%%%%%%%%%%%%%%%%%%%%%%%%%%%%%%%%%%%
%%%%%%%%%%%%%%%%%%%%%%%%%%%%%%%%%%%%%%%%%%%%%%%%%%%%%%%%%%%%%%%%%%%%%%%%%%%%%%%%%%%%%%%%%%%%%%%%%%%%%%%%%%%%%%%%%%%%%%%%%%%%
\bibliographystyle{aa}
\bibliography{biblio.bib}

%\listofobjects

%%%%%%%%%%%%%%%%%%%%%%%%%%%%%%%%%%%%%%%%%%%%%%%%%%%%%%%%%%%%%%%%%%%%%%%%%%%%%%%%%%%%%%%%%%%%%%%%%%%%%%%%%%%%%%%%%%%%%%%%%%%%
%%%%%%%%%%%%%%%%%%%%%%%%%%%%%%%%%%%%%%%%%%%%%%%%%%%%%%%%%%%%%%%%%%%%%%%%%%%%%%%%%%%%%%%%%%%%%%%%%%%%%%%%%%%%%%%%%%%%%%%%%%%%
\Online
\begin{appendix}

%%%%%%%%%%%%%%%%%%%%%%%%%%%%%%%%%%%%%%%%%%%
\section{Journal of observations}
\label{jofobs}

Table \ref{tabA1} provides the complete journal of observations.

\begin{table*}[ht]
\centering
\caption{Journal of observations.}
\begin{tabular}{lcccccc}
        	\hline
        	\hline
        	Star & Instrument & date & UT & Exp. time (sec) & S/N \\
       	\hline
        	HD~108                          & NARVAL   & Jul 21$^{\rm st}$ 2009 & 00:52:37 & 4 $\times$ 750 & 150--350 \\
                                                & NARVAL   & Jul 21$^{\rm st}$ 2009 & 02:18:22 & 4 $\times$ 750 & 200--500 \\
                                                & NARVAL   & Jul 21$^{\rm st}$ 2009 & 03:11:49 & 4 $\times$ 750 & 200--500 \\
                                                & NARVAL   & Jul 25$^{\rm th}$ 2009 & 00:41:56 & 4 $\times$ 750 & 200--500 \\
                                                & NARVAL   & Jul 25$^{\rm th}$ 2009 & 01:35:23 & 4 $\times$ 750 & 200--500 \\
                                                & NARVAL   & Jul 25$^{\rm th}$ 2009 & 02:28:49 & 4 $\times$ 750 & 200--500 \\
                                                & NARVAL   & Jul 26$^{\rm th}$ 2009 & 01:47:60 & 4 $\times$ 700 & 100--350 \\
                                                & NARVAL   & Jul 26$^{\rm th}$ 2009 & 02:38:05 & 4 $\times$ 700 & 100--400 \\
                                                & NARVAL   & Jul 26$^{\rm th}$ 2009 & 03:28:11 & 4 $\times$ 700 & 150--450 \\
                                                & NARVAL   & Jul 27$^{\rm th}$ 2009 & 00:53:23 & 4 $\times$ 806 & 100--300 \\
                                                & NARVAL   & Jul 27$^{\rm th}$ 2009 & 01:50:33 & 4 $\times$ 806 & 120--370 \\
                                                & NARVAL   & Jul 27$^{\rm th}$ 2009 & 02:47:43 & 4 $\times$ 806 & 100--350 \\
                                                & NARVAL   & Jul 28$^{\rm th}$ 2009 & 00:51:16 & 4 $\times$ 900 & 100--300 \\
                                                & NARVAL   & Jul 28$^{\rm th}$ 2009 & 01:54:42 & 4 $\times$ 900 & 100--400 \\
                                                & NARVAL   & Jul 28$^{\rm th}$ 2009 & 02:58:06 & 4 $\times$ 900 & 150--450 \\
                                                & NARVAL   & Jul 29$^{\rm th}$ 2009 & 00:39:56 & 4 $\times$ 900 & 200--550 \\
                                                & NARVAL   & Jul 29$^{\rm th}$ 2009 & 01:43:21 & 4 $\times$ 900 & 250--600 \\
                                                & NARVAL   & Jul 29$^{\rm th}$ 2009 & 02:46:46 & 4 $\times$ 900 & 200--500 \\
                                                & NARVAL   & Jul 30$^{\rm th}$ 2009 & 00:56:23 & 4 $\times$ 900 & 250--630 \\
                                                & NARVAL   & Jul 30$^{\rm th}$ 2009 & 01:59:50 & 4 $\times$ 900 & 250--600 \\
                                                & NARVAL   & Jul 30$^{\rm th}$ 2009 & 03:03:16 & 4 $\times$ 900 & 250--600 \\
                                                & NARVAL   & Jul 31$^{\rm st}$ 2009 & 00:30:28 & 4 $\times$ 812 & 200--500 \\
                                                & NARVAL   & Jul 31$^{\rm st}$ 2009 & 01:28:02 & 4 $\times$ 812 & 200--500 \\
                                                & NARVAL   & Jul 31$^{\rm st}$ 2009 & 02:25:37 & 4 $\times$ 812 & 200--550 \\
                                                & NARVAL   & Jul 31$^{\rm st}$ 2009 & 03:23:11 & 4 $\times$ 812 & 200--500 \\
                                                & NARVAL   & Aug 1$^{\rm st}$ 2009 & 00:55:57 & 4 $\times$ 812 & 200--550 \\
                                                & NARVAL   & Aug 1$^{\rm st}$ 2009 & 01:57:06 & 4 $\times$ 812 & 250--600 \\
                                                & NARVAL   & Aug 1$^{\rm st}$ 2009 & 02:54:39 & 4 $\times$ 812 & 200--470 \\
                                                & NARVAL   & Aug 2$^{\rm nd}$ 2009 & 00:50:14 & 4 $\times$ 900 & 200--470 \\
                                                & NARVAL   & Aug 2$^{\rm nd}$ 2009 & 01:53:47 & 4 $\times$ 900 & 150--400 \\
                                                & NARVAL   & Aug 2$^{\rm nd}$ 2009 & 02:57:24 & 4 $\times$ 900 & 150--400 \\
                                                & NARVAL   & Aug 4$^{\rm th}$ 2009 & 01:24:42 & 4 $\times$ 900 & 200--470 \\
                                                & NARVAL   & Aug 4$^{\rm th}$ 2009 & 02:28:07 & 4 $\times$ 900 & 250--600 \\
                                                & NARVAL   & Aug 4$^{\rm th}$ 2009 & 03:31:34 & 4 $\times$ 900 & 200--570 \\
                                                & NARVAL   & Aug 5$^{\rm th}$ 2009 & 00:22:43 & 4 $\times$ 900 & 200--500 \\
                                                & NARVAL   & Aug 5$^{\rm th}$ 2009 & 01:26:09 & 4 $\times$ 900 & 200--600 \\
                                                & NARVAL   & Aug 5$^{\rm th}$ 2009 & 02:29:35 & 4 $\times$ 900 & 200--600 \\
        	\hline
	        HD~148937			&  ESPADONS & Jun 20$^{\rm th}$ 2010 & 09:10:53 & 4 $\times$ 600 & 300--850 & \\
                                               	&  ESPADONS & Jun 21$^{\rm st}$ 2010 & 08:31:20 & 4 $\times$ 600 & 250--750 & \\
                                               	&  ESPADONS & Jun 22$^{\rm nd}$ 2010 & 09: 9:60 & 4$\times$ 600 & 170--550 & \\
                                               	&  ESPADONS & Jun 23$^{\rm rd}$ 2010 & 08:11:34 & 4 $\times$ 600 & 150--400 & \\
                                               	&  ESPADONS & Jun 24$^{\rm th}$ 2010 & 08: 8:21 & 4 $\times$ 600 & 300--800 & \\
                                               	&  ESPADONS & Jun 25$^{\rm th}$ 2010 & 08:16:39 & 4 $\times$ 600 & 3250--800 & \\
	      \hline
        	HD~191612               	& ESPADONS & Aug 23$^{\rm rd}$ 2008 & 09:43:36 & 4 $\times$ 300 & 100--300 \\
        				       	& ESPADONS & Aug 23$^{\rm rd}$ 2008 & 10:04:20 & 4 $\times$ 300 & 100--350 \\
        				       	& ESPADONS & Aug 23$^{\rm rd}$ 2008 & 10:25:05 & 4 $\times$ 300 & 100--350 \\	
				       		& ESPADONS & Aug 23$^{\rm rd}$ 2008 & 10:45:49 & 4 $\times$ 300 & 100--350 \\
        				       	& ESPADONS & Aug 23$^{\rm rd}$ 2008 & 11:07:00 & 4 $\times$ 300 & 100--350 \\
        				       	& ESPADONS & Aug 23$^{\rm rd}$ 2008 & 11:27:45 & 4 $\times$ 300 & 100--350 \\	
				       		& ESPADONS & Aug 23$^{\rm rd}$ 2008 & 11:48:29 & 4 $\times$ 300 & 100--350 \\
       		                                & ESPADONS & Aug 23$^{\rm rd}$ 2008 & 12:09:13 & 4 $\times$ 300 & 100--350 \\	
        	\hline
\end{tabular}
\label{tabA1}
\end{table*}

\setcounter{table}{0}

\begin{table*}[ht]
\centering
\caption{Continued.}
\begin{tabular}{lcccccc}
        	\hline
        	Star & Instrument & date & UT & Exp. time (sec) & S/N \\
       	\hline
        	$\Theta^{1}$ Ori C 		& NARVAL & Mar 10$^{\rm th}$ 2007 & 08:28:19 & 4 $\times$ 400 & 500--1400 \\
                                                & NARVAL & Mar 10$^{\rm th}$ 2007 & 08:59:01 & 4 $\times$ 400 & 450--1250 \\
        	\hline
        	$\zeta$ Ori A 			& NARVAL & Oct 18$^{\rm th}$ 2007 & 00:52:56 & 48 $\times$ 4 $\times$ 20 & 780 -- 990 \\
			      			& NARVAL & Oct 19$^{\rm th}$ 2007 & 04:35:04 &   8 $\times$ 4 $\times$ 40 & 1010 -- 1080 \\
			      			& NARVAL & Oct 20$^{\rm th}$ 2007 & 00:59:39 & 44 $\times$ 4 $\times$ 40 & 1220 -- 1470 \\
			      			& NARVAL & Oct 21$^{\rm st}$ 2007 & 23:06:15 & 48 $\times$ 4 $\times$ 40 & 810 -- 1460 \\
			      			& NARVAL & Oct 22$^{\rm nd}$ 2007 & 23:45:52 & 48 $\times$ 4 $\times$ 40 & 1090 -- 1480 \\
			      			& NARVAL & Oct 23$^{\rm rd}$ 2007 & 23:13:45 & 48 $\times$ 4 $\times$ 40 & 1030 -- 1480 \\
			      			& NARVAL & Oct 24$^{\rm th}$ 2007 & 23:58:34 & 48 $\times$ 4 $\times$ 40 & 1200 -- 1470 \\
        	\hline
        	HD~57682 	                & ESPADONS & May 4$^{\rm th}$ 2009 & 06:05:55 & 4 $\times$ 600 & 500--1130 \\
						& ESPADONS & May 5$^{\rm th}$ 2009 & 06:22:17 & 4 $\times$ 540 & 450--1000 \\
						& ESPADONS & May 7$^{\rm th}$ 2009 & 06:46:41 & 4 $\times$ 540 & 350--700 \\
						& ESPADONS & May 8$^{\rm th}$ 2009 & 06:21:09 & 4 $\times$ 540 & 350--800 \\
						& ESPADONS & May 9$^{\rm th}$ 2009 & 06:20:36 & 4 $\times$ 540 & 300--600 \\
        	\hline
        	$\tau$ Sco 	                & ESPADONS & May 23$^{\rm rd}$ 2005 & 09:14:25 & 4 $\times$ 60 & 700--1550 \\
                                                & ESPADONS & Sep 20$^{\rm th}$ 2005 & 05:01:56 & 4 $\times$ 30 & 600--1200 \\
                                                & ESPADONS & Aug 5$^{\rm th}$ 2006 & 05:25:22 & 4 $\times$ 30 & 500--1150 \\
                                                & ESPADONS & Mar 4$^{\rm th}$ 2007 & 16:23:15 & 4 $\times$ 30 & 700--1350 \\
                                                & ESPADONS & Jun 29$^{\rm th}$ 2008 & 08:03:07 & 4 $\times$ 45 & 400--1050 \\
        	\hline
\end{tabular}
\label{tabA1}
\end{table*}

\end{appendix}

%%%%%%%%%%%%%%%%%%%%%%%%%%%%%%%%%%%%%%%%%%%%%%%%%%%%%%%%%%%%%%%%%%%%%%%%%%%%%%%%%%%%%%%%%%%%%%%%%%%%%%%%%%%%%%%%%%%%%%%%%%%%
%%%%%%%%%%%%%%%%%%%%%%%%%%%%%%%%%%%%%%%%%%%%%%%%%%%%%%%%%%%%%%%%%%%%%%%%%%%%%%%%%%%%%%%%%%%%%%%%%%%%%%%%%%%%%%%%%%%%%%%%%%%%
\end{document}